\documentclass[journal]{IEEEtran}

\ifCLASSINFOpdf
\else
   \usepackage[dvips]{graphicx}
\fi
\usepackage{url}

\usepackage{graphicx}
\usepackage{amssymb}
\usepackage{textcomp}
\usepackage{stfloats}
\usepackage{url}
\usepackage{verbatim}
\usepackage{graphicx}
\usepackage{cite}
\usepackage{amsmath,amsfonts}
\usepackage{algorithmic}
\usepackage{algorithm}
\usepackage{array}
\usepackage[colorlinks,linkcolor=red]{hyperref}
\usepackage{booktabs}
\usepackage[caption=false]{subfig}

\usepackage{makecell}
\usepackage{multirow}

\begin{document}
\title{Graph Attention for Automated Audio Captioning}
\author{Feiyang Xiao, Jian Guan, \IEEEmembership{Member, IEEE}, Qiaoxi Zhu, \IEEEmembership{Member, IEEE}, \\ and Wenwu Wang, \IEEEmembership{Senior Member, IEEE} 
\thanks{
	This work was partly supported by the Natural Science Foundation of Heilongjiang Province under Grant No. YQ2020F010, a Newton Institutional Links Award from the British Council with Grant No. 623805725, and a GHfund B with Grant No. 202302026860. (Corresponding author: Jian Guan)
}
\thanks{F. Xiao and J. Guan are with the Group of Intelligent Signal Processing (GISP), College of Computer Science and Technology, Harbin Engineering University, Harbin 150001, China (emails: xiaofeiyang128@gmail.com; j.guan@hrbeu.edu.cn).}
\thanks{Q. Zhu is with the Centre for Audio, Acoustics and Vibration, University of Technology Sydney, Ultimo, NSW, Australia (email: qiaoxi.zhu@gmail.com).}
\thanks{W. Wang is with the Centre for Vision, Speech and Signal Processing, University of Surrey, Guildford GU2 7XH, U.K.  (email: w.wang@surrey.ac.uk).}}

\maketitle

\begin{abstract}
 State-of-the-art audio captioning methods typically use the encoder-decoder structure with pretrained audio neural networks (PANNs) as encoders for feature extraction. However, the convolution operation used in PANNs is limited in capturing the long-time dependencies within an audio signal, thereby leading to potential performance degradation in audio captioning. This letter presents a novel method using graph attention (GraphAC) for encoder-decoder based audio captioning. In the encoder, a graph attention module is introduced after the PANNs to learn contextual association (i.e. the dependency among the audio features over different time frames) through an adjacency graph, and a top-{\it{k}} mask is used to mitigate the interference from noisy nodes. The learnt contextual association leads to a more effective feature representation with feature node aggregation. As a result, the decoder can predict important semantic information about the acoustic scene and events based on the contextual associations learned from the audio signal. Experimental results show that GraphAC outperforms the state-of-the-art methods with PANNs as the encoders, thanks to the incorporation of the graph attention module into the encoder for capturing the long-time dependencies within the audio signal. The source code is available at \url{https://github.com/LittleFlyingSheep/GraphAC}.

\end{abstract}


\IEEEpeerreviewmaketitle

\section{Introduction}
\label{sec:1}

\IEEEPARstart{A}{utomated} audio captioning (AAC) aims to describe an audio signal with captions using natural language and focus on non-speech content, such as environmental sound \cite{drossos2020clotho}. It can facilitate man-machine interaction for those with hearing loss, sound analysis for security surveillance \cite{xinhao2021_t6}, and automatic content summarisation, e.g., subtitling for the sound of a television program \cite{xinhao2021_t6, xu2021_t6}.

The encoder-decoder structure is popular for AAC. The audio encoder extracts the audio feature, and the text decoder generates the caption from the audio feature. In early methods \cite{drossos2017automated, Cakir2020, Koizumi2020}, recurrent neural networks (RNNs) \cite{schuster1997bidirectional} and Transformer \cite{NIPS2017_3f5ee243} have been used for audio captioning. However, the encoders used in these methods may not be effective in feature representation, due to the use of either a simple model or the limited amount of training data. As a solution, the pretrained audio neural networks (PANNs) \cite{kong2020panns} was applied widely as the audio encoder in recent research. The PANNs model is pretrained on a large-scale audio dataset (i.e., AudioSet \cite{gemmeke2017audio}).
This has led to significant performance improvement in audio captioning \cite{xinhao2021_t6, narisetty2021_t6, chen2021_t6, xu2021_t6, Ye2021, Han2021}. 

Despite its excellent performance, the convolution operation in PANNs primarily captures information from the local receptive field (i.e., local time-frequency region), ignoring the contextual associations among audio features and their long-time dependencies \cite{mei2022automated, 9626590}. Nevertheless, audio signals are time-variant and contain rich temporal information, including long-time dependencies that carry semantic information about the acoustic scene and events. Missing such information may affect the effectiveness of audio representation in the audio encoder and limit the captioning performance. 

This letter presents a novel audio captioning (AC) method, namely GraphAC,  with a graph attention module incorporated into the encoder for feature representation. Specifically, in GraphAC, P-Transformer \cite{mei2022automated} is used as the backbone, and the graph attention module is introduced after the PANNs in the audio encoder to obtain more effective audio representation. The graph attention module not only captures the temporal contextual information within the audio signal, i.e., by exploiting the contextual association between audio nodes (i.e., audio feature frames) obtained in the learnt adjacency graph with a top-{\it{k}} mask, but also highlights the important semantic information about the acoustic scene and events in the feature representation, i.e., by aggregating audio nodes with the learnt adjacency graph. As a result, the encoder of the proposed GraphAC acquires a better audio feature due to the exploitation of the contextual information from the longer time duration. This information can improve the accuracy of captions generated by the text decoder (i.e., a Transformer-based decoder).

The graph attention module learns the edge connections between audio feature nodes via the attention mechanism \cite{velivckovic2018graph}, and differs significantly from the graph convolutional network (GCN), which is popular for image and video captioning \cite{GCN_image, GCN_video}, but uses convolution as the fundamental operation for feature representation. In contrast to GCN for image and video captioning, our GraphAC does not require a pre-trained graph model to generate a graph structure. In addition, it introduces a top-{\it{k}} mask strategy to remove noisy nodes caused by non-zero weights assigned to audio frames in encoder learning \cite{mei2022automated}. Compared to RNNs and Transformers, as used in official baselines of DCASE Challenge Task 6, which can also model long-time dependencies, our method offers an additional advantage in attending important audio feature nodes and ignoring meaningless audio feature nodes, thereby highlighting the important semantic information about acoustic scenes and events.

Experiments are performed on the DCASE 2021 Challenge Task 6 dataset~\cite{drossos2017automated}, and show that the proposed GraphAC outperforms state-of-the-art techniques in all the assessment criteria by exploiting time dependencies. With the graph attention based audio feature representation, the proposed method can capture important semantic information about the acoustic scene and events which helps improve the captioning performance. 

\section{Graph Attention Audio Captioning}
\label{sec:GraphAC}

\begin{figure*}[htbp]
    \centering
    \centerline{\includegraphics[width=0.83\textwidth]{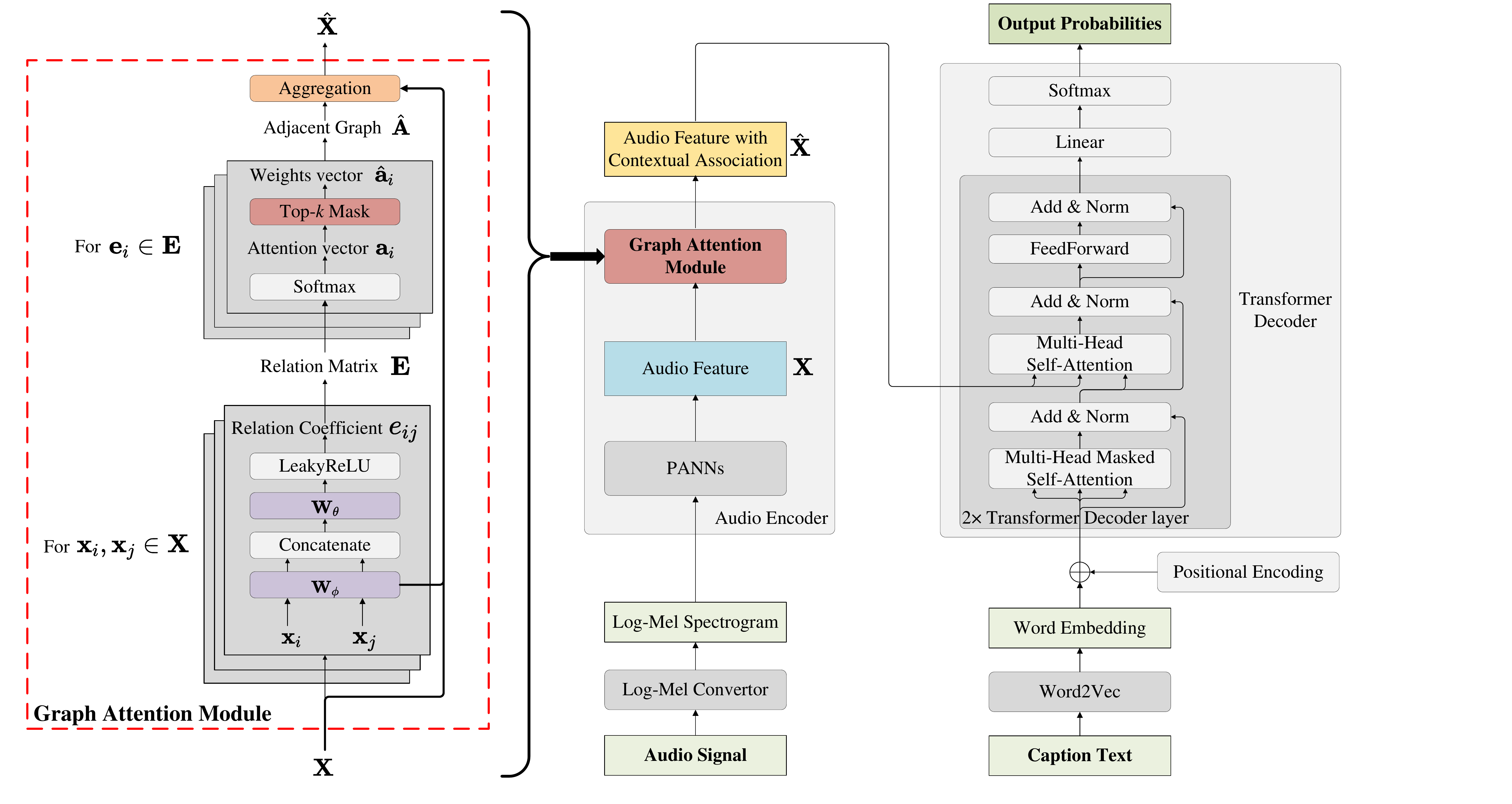}}
    \vspace{-3mm}
    \caption{The framework of our proposed GraphAC method, where P-Transformer \cite{xinhao2021_t6} is used as the backbone. The difference between GraphAC and P-Transformer is that GraphAC has the graph attention module in the encoder, as shown in the red dashed box.}
    \vspace{-4mm}
    \label{fig:framework}
\end{figure*}

The proposed GraphAC adopts P-Transformer \cite{xinhao2021_t6} as the backbone, with the graph attention module added in the encoder to model the relations among the audio nodes (i.e., audio feature frames) extracted by PANNs, and to obtain an improved audio feature representation. Then, a Transformer decoder is employed to predict the caption from the audio feature representation. Fig. \ref{fig:framework} shows the framework of GraphAC.

\subsection{Audio Feature Extraction}

The PANNs module \cite{kong2020panns} is applied to extract the audio feature from an audio signal. The audio signal is converted to the log-Mel spectrogram $\mathbf{X}_{Mel}$ as input to the PANNs module (i.e., CNN10). Different from the original CNN10 structure in \cite{kong2020panns}, here, only the global average pooling is used on the Mel-band dimension after the convolutional blocks, and the channel dimension is taken as the audio feature dimension in this work. Then, the dimension of the output of the last two layers is modified to obtain the audio feature $\mathbf{X} \in \mathbb{R}^{T \times D}$, where $T$ denotes the temporal dimension and $D$ denotes the audio feature dimension. Here, $D$ is set empirically as 128.

\subsection{Graph Attention for Audio Feature Representation}

Instead of directly using the audio feature extracted by PANNs, a graph attention module is introduced to represent the timing information of the audio signal. The adjacency graph is built in the encoder to represent the audio feature with node relations by graph attention mechanism, where a top-{\it{k}} mask strategy is introduced to remove noisy nodes for better feature representation. Then, the audio feature nodes are aggregated with the learnt contextual association.

\subsubsection{\textbf{Audio Feature Graph Modelling}}

The contextual relation between audio feature nodes is built by the adjacency graph to exploit the long-time dependencies within the audio feature. The audio feature $\mathbf{X} \!=\! [\mathbf{x}_1, \mathbf{x}_2, \cdots, \mathbf{x}_T]^{\top}$ is a set of $T$ audio feature nodes $\mathbf{x}_i \in \mathbb{R}^{D}$ ($1 \le i \le T$) and $\top$ denotes the matrix transposition. Each node represents the audio feature over a time frame. The coefficient characterising the relation between two audio feature nodes $\mathbf{x}_i$ and $\mathbf{x}_j$ ($1 \le i, j \le T$) is calculated by a learnable linear mapping operation via the additive score function following \cite{velivckovic2018graph}: 
\begin{equation}
\label{eq:relation_coefficient}
    e_{ij} = LeakyReLU(\mathbf{W}_{\theta}[\mathbf{W}_{\phi}\mathbf{x}_i; \mathbf{W}_{\phi}\mathbf{x}_j]),
\end{equation}
where $\mathbf{W}_{\phi} \in \mathbb{R}^{D \times D}$ and $\mathbf{W}_{\theta} \in \mathbb{R}^{1 \times 2D}$ are two matrices containing learnable parameters, and $[\cdot  ; \cdot  ]$ denotes the concatenation operation. The matrix $\mathbf{W}_{\phi}$ maps each audio feature node into a relation embedding space, and the matrix $\mathbf{W}_{\theta}$ maps the concatenated relation embedding into the relation coefficient $e_{ij}$. The leaky ReLU is used as the activation function. The relation matrix which is denoted as $\mathbf{E} \in \mathbb{R}^{T \times T}$ with $e_{ij}$ being its element at the $i$-th row and $j$-th column, contains the relation coefficients of all pairs of nodes. 

Then, we employ the relation matrix to calculate the adjacency graph. The row vector of the relation matrix $\mathbf{e}_i$ ($1\!\le\!i\!\le\!T$)
is normalised by the softmax function, resulting in the attention vector $\mathbf{a}_i$. Here, $\mathbf{a}_i$ contains the edge weights between the node $\mathbf{x}_i$ and all nodes, including itself. Then, we adopt a top-$k$ mask strategy for node selection as follows
\begin{equation}
\label{eq:topk}
    \hat{a}_{ij} = \left\{\begin{matrix}
                            a_{ij},  & a_{ij} \in {\text{top}_{\textit{k}}}(\mathbf{a}_i)\\
                            \ 0,     & a_{ij} \notin {\text{top}_{\textit{k}}}(\mathbf{a}_i)
                          \end{matrix}\right.,
\end{equation}
where $a_{ij}$ denotes the input weight between the node $\mathbf{x}_i$ and the node $\mathbf{x}_j$, and ${\text{top}_{\textit{k}}}(\mathbf{a}_i)$ denotes the set of $\textit{k}$ largest elements in $\mathbf{a}_i$. {With the top-{\it{k}} mask, we can prioritise important relations between audio feature nodes and select $\textit{k}$ most relevant nodes, while mitigating the interferences from noisy nodes.}
Finally, the adjacency graph $\hat{\mathbf{A}} \in \mathbb{R}^{T \times T}$ is formed with $\hat{a}_{ij}$ being its $ij$-th element.

\subsubsection{\textbf{Graph Nodes Aggregation}}

With the learnt adjacency graph, we aggregate audio feature nodes to obtain the audio feature $\hat{\mathbf{X}}$ with the contextual association
\begin{equation}
\label{eq:aggregation}
    \hat{\mathbf{X}} = \hat{\mathbf{A}}\mathbf{X}\mathbf{W}_{\phi}^{\top} + \mathbf{X}.
\end{equation}
Here, $\hat{\mathbf{X}}$ is the output audio feature of the audio encoder, which contains the timing information of the audio signal because the learnt relations of the aggregated audio feature nodes reflect the contextual association within the audio signal.

The node aggregation with the learnt adjacency graph can highlight the important semantic information about acoustic scenes and events with the time dependency of the audio signal. With the residual connection, the obtained graph audio feature represents long-time dependency information from graph attention and local-dependency information from PANNs with external knowledge via model pretraining.

\subsection{Transformer Decoder}

To generate captions, we use the Transformer decoder which takes the audio feature with the contextual association. The decoder has two inputs. One is the audio feature with contextual association $\hat{\mathbf{X}}$, the other is the word embedding with positional encoding $\mathbf{Y} = [\mathbf{y}_0, \mathbf{y}_1, \cdots, \mathbf{y}_{N-1}]^{\top} \in \mathbb{R}^{N \times D}$, where $N$ denotes the number of words in the caption. Noting that, $\mathbf{y}_0$ is from the token $<\!\text{sos}\!>$ representing the start of the sequence, and it does not belong to the caption. The $n$-th word $\mathbf{y}_n$ in the caption is generated as
\begin{equation}
    p(\mathbf{y}_n | \hat{\mathbf{X}}, \mathbf{Y}_{pre}) = \text{Decoder}(\hat{\mathbf{X}}, \mathbf{Y}_{pre}),
\end{equation}
where $p(\mathbf{y}_n | \hat{\mathbf{X}}, \mathbf{Y}_{pre})$ is the posterior probability of the $n$-th word $\mathbf{y}_n$ and $1\le n \le N$. $\mathbf{Y}_{pre}=[\mathbf{y}_0, \cdots, \mathbf{y}_{n-1}]^{\top}$ denotes the previously generated caption words. 

We pretrain a Word2Vec language model \cite{mikolov2013efficient} by combining the captions on the Clotho-v2 dataset \cite{drossos2020clotho} and an external dataset, i.e., AudioCaps \cite{kim2019audiocaps}, for an effective word embedding. Moreover, the proposed GraphAC method is firstly pretrained on the AudioCaps dataset and then fine-tuned on the Clotho-v2 dataset, with the same training strategy as in \cite{xinhao2021_t6, xiao2022local}.

\section{Experiments and Results}
\label{sec:3}
\subsection{Dataset}
Following Task 6 of the DCASE 2021 Challenge and related work, such as \cite{xinhao2021_t6}, we use the development and validation splits of the AudioCaps dataset \cite{kim2019audiocaps} for pretraining, the development and validation splits of the Clotho-v2 dataset \cite{drossos2020clotho} for fine-tuning, and the evaluation split of the Clotho-v2 dataset for evaluation. Specifically, AudioCaps has 44366, 458, and 905 audio clips in the development, validation, and evaluation sets, and Clotho-v2 has 3839, 1045 and 1045 audio clips in the development, validation and evaluation splits, respectively.

\subsection{Experimental Setup}
\label{subsec:setup}

In the graph attention module, $\textit{k}$ was empirically set as 25 in the top-{\it{k}} mask. Following P-Transformer \cite{xinhao2021_t6}, GraphAC applied SpecAugment and mix-up strategies to improve generalisation. The cross-entropy loss with label smoothing \cite{szegedy2016rethinking} was used with Adam optimizer \cite{kingma2014adam} to optimize the network. The batch size was 16, and the learning rate was 0.0001. The decoder used a teacher forcing strategy in training and a beam search strategy with a beam size of 5 in evaluation. 

Following DCASE Challenge, all the methods are evaluated by machine translation metrics (i.e., BLEU$_n$, ROUGE$_l$ and METEOR) and captioning metrics (CIDE$_r$, SPICE and SPIDE$_r$). BLEU$_n$ \cite{papineni2002bleu} measures a modified n-gram precision. ROUGE$_l$ \cite{lin2004rouge} is a score based on the longest common sub-sequence. METEOR \cite{lavie2007meteor} is a harmonic mean of weighted unigram precision and recall. CIDE$_r$ \cite{vedantam2015cider} is a weighted cosine similarity of n-grams. SPICE \cite{anderson2016spice} is the F-score of semantic propositions extracted from caption and reference. SPIDE$_r$ \cite{liu2017improved} is the mean score between CIDE$_r$ and SPICE, which evaluates both the fluency and semantic properties of the caption. The source code\footnote{\url{https://github.com/LittleFlyingSheep/GraphAC} } along with examples of the predicted captions is released for reproducibility of our work.

\subsection{Performance Comparison}
\begin{table*}[t]
  \centering
  \caption{Performance comparison on the evaluation split of the Clotho-v2 dataset.}
  \resizebox{\textwidth}{!}
  {
    \begin{tabular}{cccccccccc}
    \toprule
    Method & \multicolumn{1}{l}{BLEU$_1$(\%)} & \multicolumn{1}{l}{BLEU$_2$(\%)} & \multicolumn{1}{l}{BLEU$_3$(\%)} & \multicolumn{1}{l}{BLEU$_4$(\%)} & \multicolumn{1}{l}{ROUGE$_l$(\%)} & \multicolumn{1}{l}{METEOR(\%)} & \multicolumn{1}{l}{CIDE$_r$(\%)} & \multicolumn{1}{l}{SPICE(\%)} & \multicolumn{1}{l}{SPIDE$_r$(\%)} \\
    \midrule
    SJTU \cite{xu2021_t6} & 56.5 & - & - & 15.5 & 37.4 & 17.4 & 39.9 & 11.9 & 25.9 \\
    P-Conformer \cite{narisetty2021_t6} & 54.1 & 34.6 & 23.1 & 15.2 & 35.6 & 16.1 & 36.2 & 11.0 & 23.6 \\
    CNN14-M2Transformer \cite{chen2021_t6} & 55.5 & 35.7 & 23.6 & 15.3 & 36.6 & 16.8 & 40.9 & 12.0 & 26.5 \\
    MAAC \cite{Ye2021} & 57.7 & - & - & 17.4 & 37.7 & 17.4 & 41.9 & 11.9 & 26.9  \\
    EaseAC \cite{Han2021} & 55.4 & 35.6 & 23.5 & 15.3 & 36.4 & 16.7 & 40.5 & 11.7 & 26.1 \\
    \midrule
    P-Transformer (backbone) \cite{xinhao2021_t6} & 56.1 & 37.4 & 25.7 & 17.4 & 37.9 & 17.1 & 42.6 & 12.4 & 27.5 \\
    \textbf{GraphAC w/o top-{\it{k}}} & 58.0 & \textbf{38.8} & \textbf{26.5} & 17.7 & 38.4 & \textbf{17.8} & 43.5 & 12.4 & 27.9 \\
    \textbf{GraphAC} & \textbf{58.1} & 38.6 & \textbf{26.5} & \textbf{18.1} & \textbf{38.5} & 17.5 & \textbf{43.7} & \textbf{12.6} & \textbf{28.1} \\
    \bottomrule
    \end{tabular}
  }
  \label{tab:performance}
  \vspace{-2mm}
\end{table*}
We compare the proposed method with the state-of-the-art methods that all use PANNs as the encoder to extract the audio feature but do not model the long-time dependencies, including P-Transformer \cite{xinhao2021_t6} (backbone method), SJTU \cite{xu2021_t6}, P-Conformer \cite{narisetty2021_t6}, CNN14-M2Transformer \cite{chen2021_t6}, MAAC \cite{Ye2021} and EaseAC \cite{Han2021}. All these methods adopt Word2Vec in the decoder to obtain the word embedding for caption prediction,  except EaseAC and P-Conformer.  Since reinforcement learning \cite{reinforce} is not employed in the proposed GraphAC, for fair comparisons, it is not used in any of the compared methods in our experiments. Note that, without reinforcement learning does not affect the main conclusion drawn in the comparisons. For a fair comparison, EaseAC is pretrained on the AudioCaps dataset without using the private dataset in \cite{Han2021}. 

Table \ref{tab:performance} shows the performances of the proposed GraphAC method and the state-of-the-art methods. The proposed GraphAC outperforms these state-of-the-art methods in all evaluation metrics, including SPIDE$_r$, the most important caption metric in the ranking of the DCASE Challenge. Different from other methods, the proposed GraphAC method models the long time dependencies of the audio feature through graph attention. The result shows the effectiveness of the proposed GraphAC method and the importance of modelling long-range temporal information in audio feature representation for the audio captioning task. Without the graph attention module, GraphAC is reduced to the backbone method (i.e., P-Transformer).

\subsection{Effect of Graph Attention on Audio Feature Representation}
\begin{figure}[tbp]
    \vspace{-4mm}
	\centering
	\subfloat[\centering{Spectrogram}]{
	    \label{subfig:spectrogram_1}
	    \includegraphics[width=0.45\linewidth]{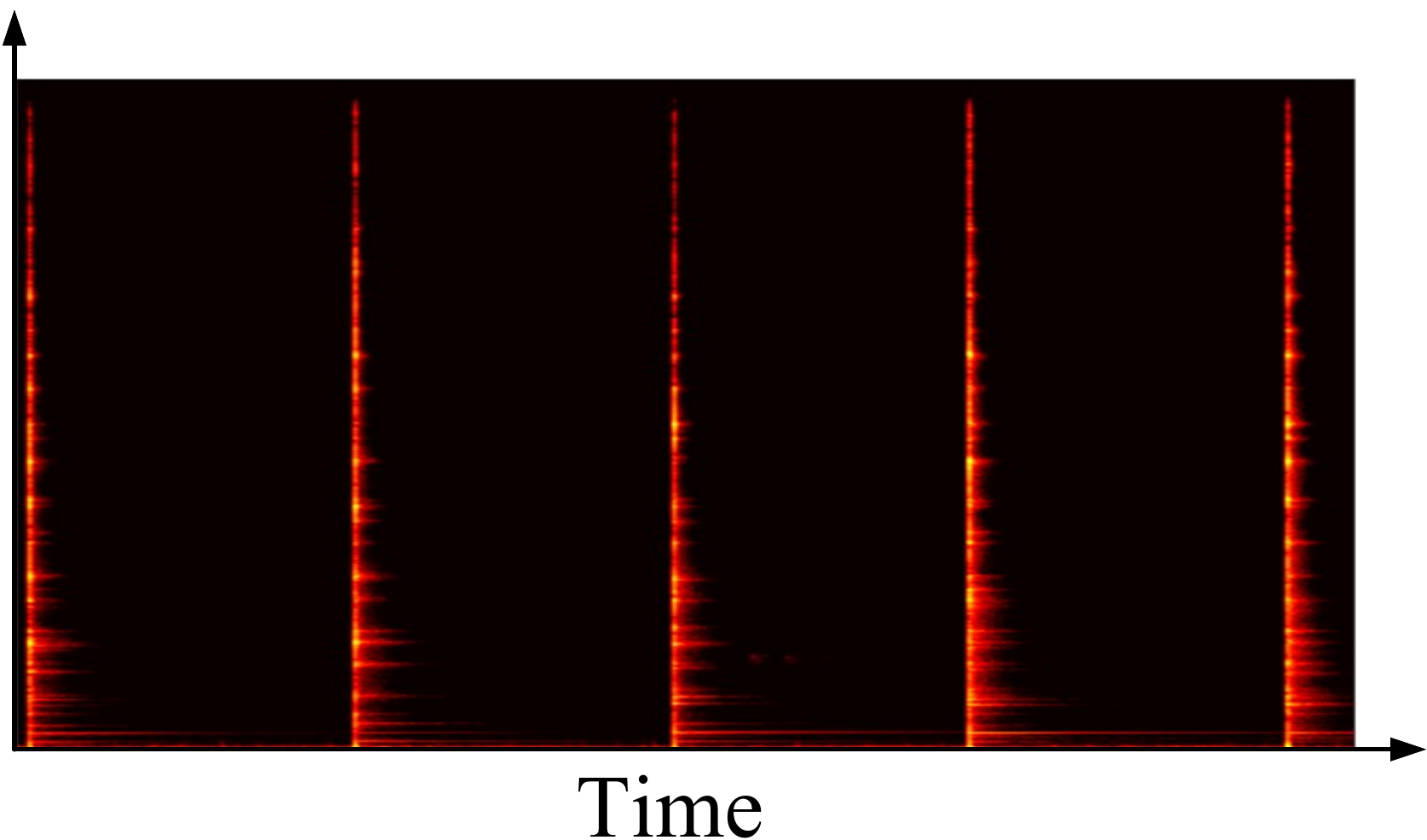}
	    }
	\quad\vspace{-4mm}
	\subfloat[\centering{Spectrogram}]{
	    \label{subfig:spectrogram_2}
	    \includegraphics[width=0.45\linewidth]{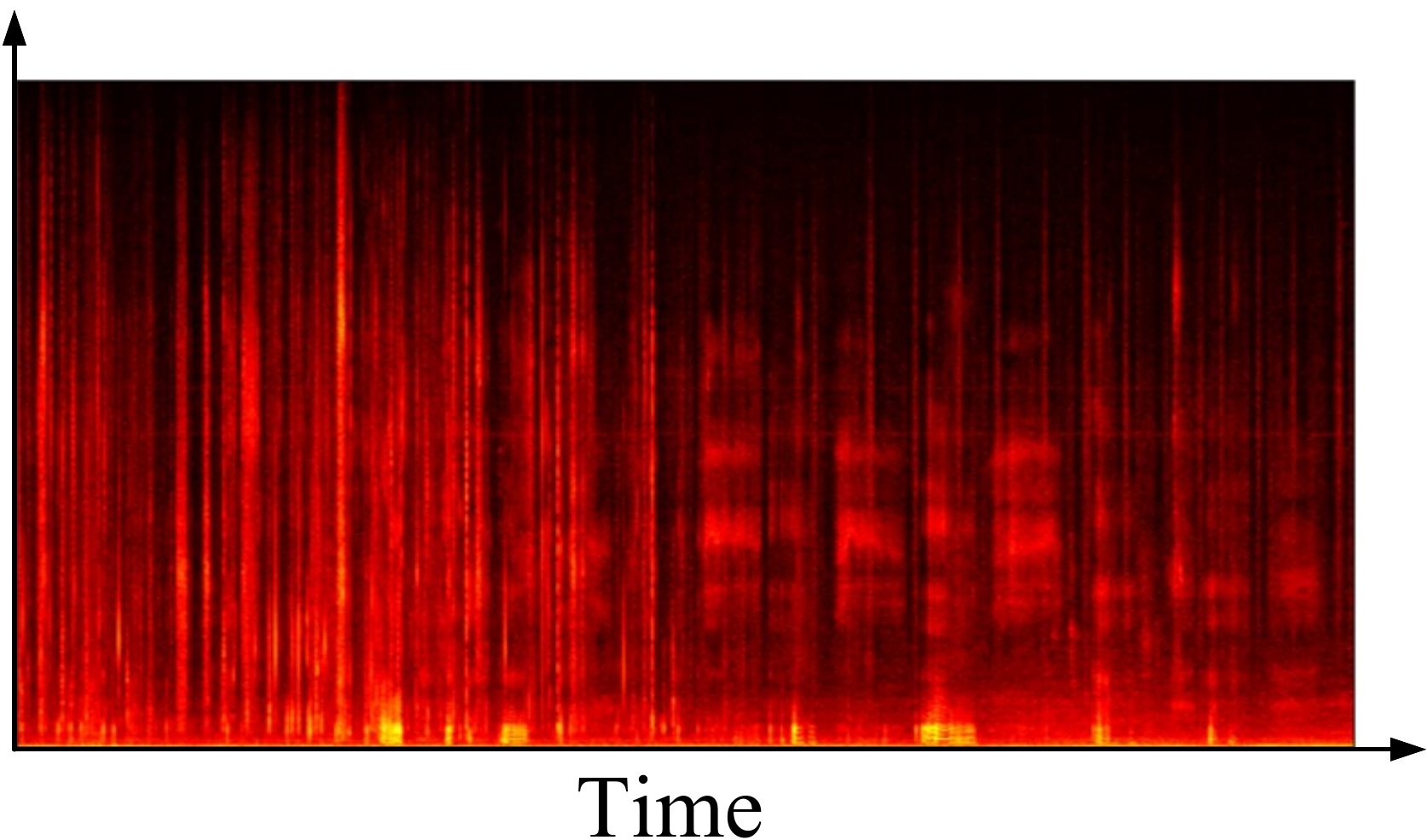}
	    }
	\quad\vspace{-4mm}
	\subfloat[\centering{Interpolated adjacency graph $\hat{\mathbf{A}}$ in GraphAC}]{
    	\label{subfig:graph_1}
    	\includegraphics[width=0.45\linewidth]{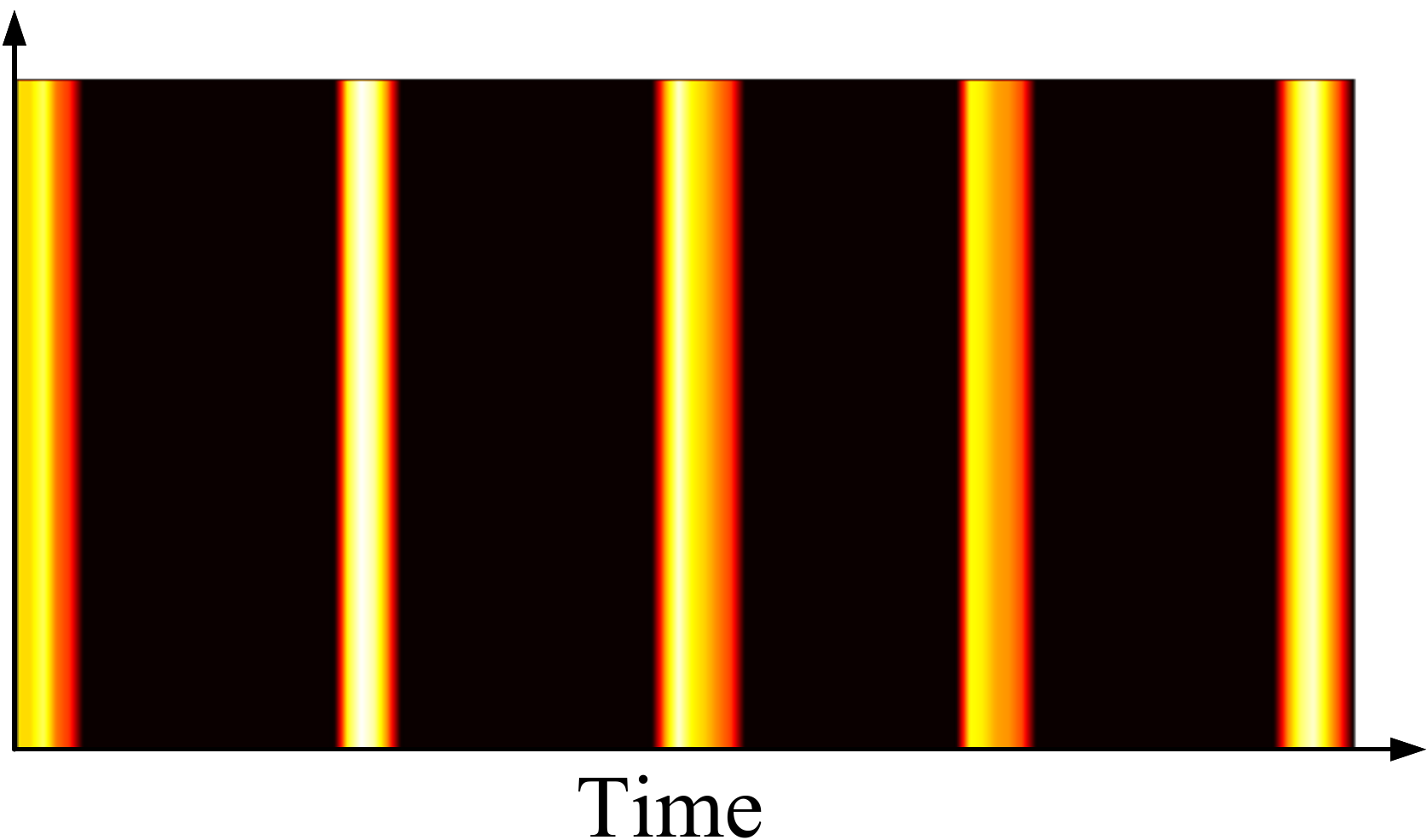}
        }
    \quad
    \subfloat[\centering{Interpolated adjacency graph $\hat{\mathbf{A}}$ in GraphAC}]{
    	\label{subfig:graph_2}
    	\includegraphics[width=0.45\linewidth]{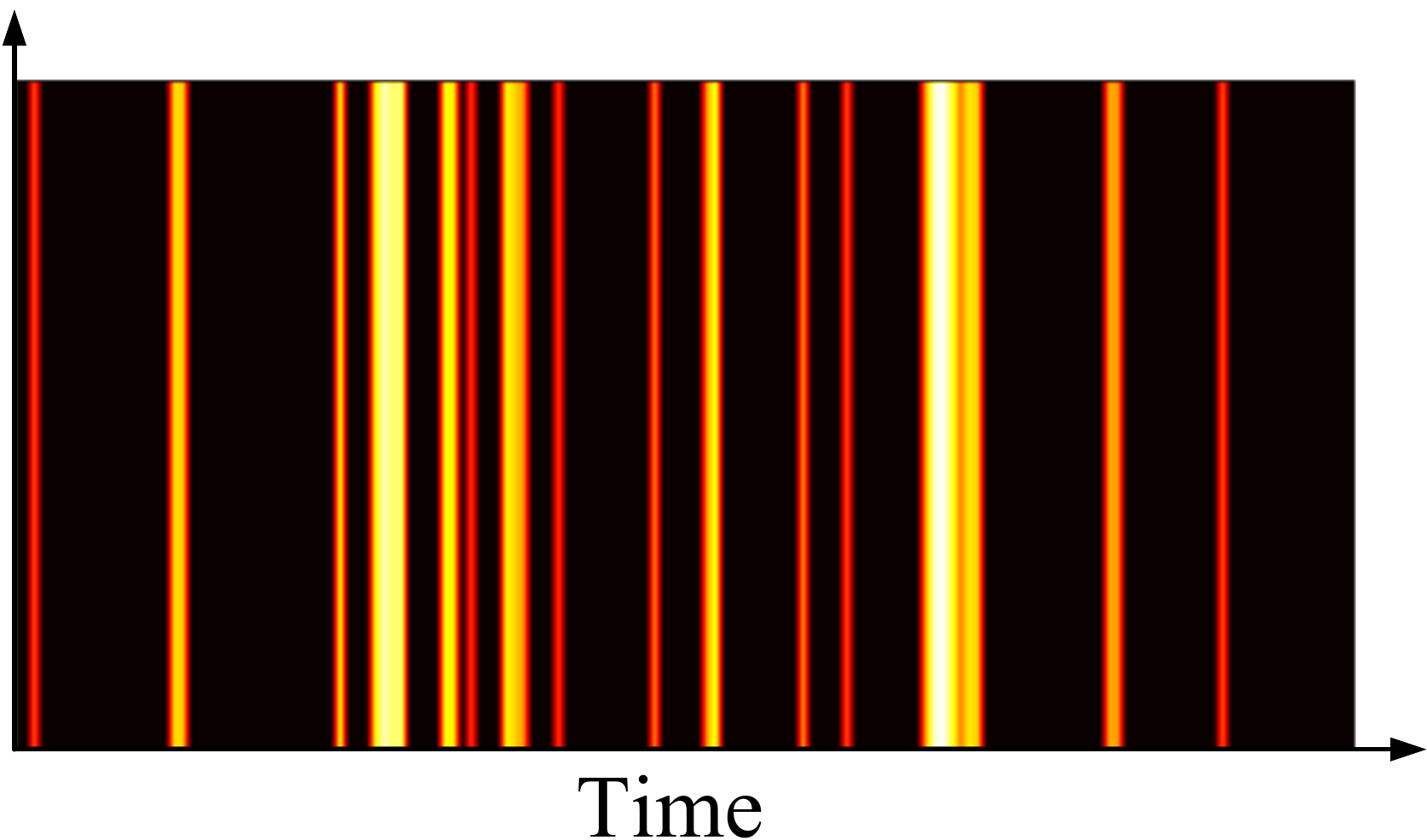}
        }
    \quad
    \subfloat[\centering{Interpolated adjacency graph $\hat{\mathbf{A}}$ in GraphAC w/o top-{\it{k}}}]{
	    \label{subfig:without_topk_1}
	    \includegraphics[width=0.45\linewidth]{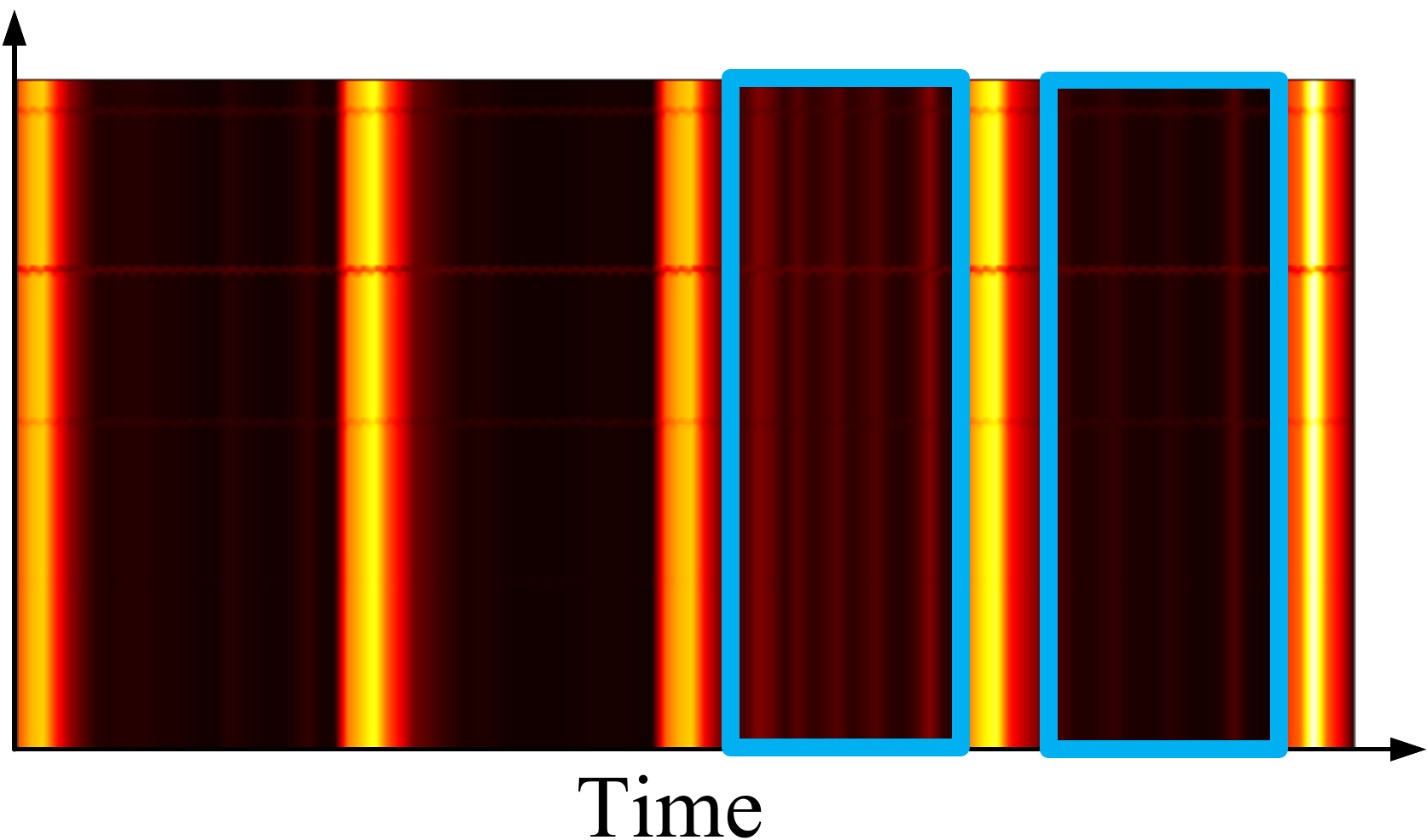}
	    }
	\quad
	\subfloat[\centering{Interpolated adjacency graph $\hat{\mathbf{A}}$ in GraphAC w/o top-{\it{k}}}]{
	    \label{subfig:without_topk_2}
	    \includegraphics[width=0.45\linewidth]{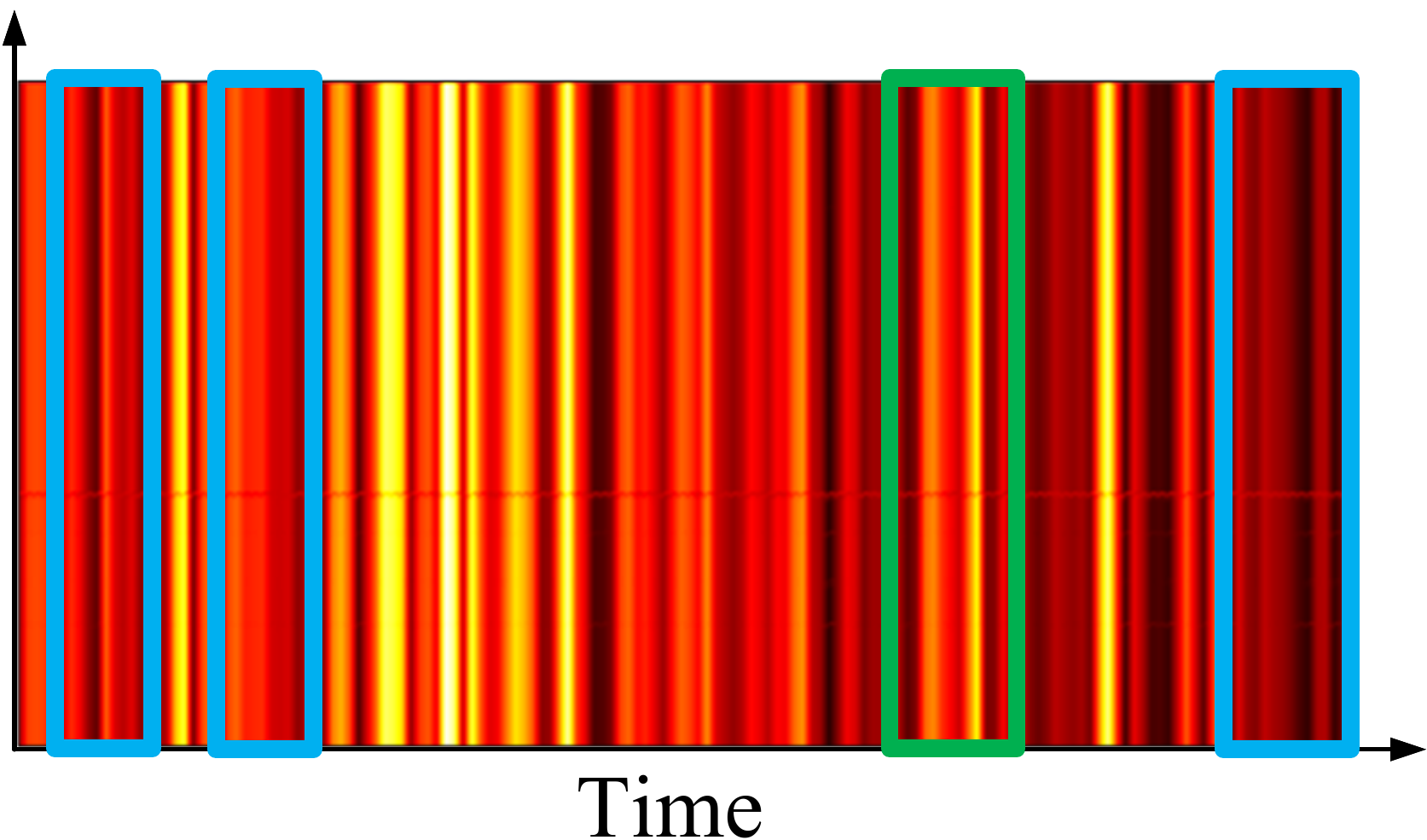}
	    }
\caption{
	Audio feature representation. The left column is a discrete sound ``five different sounding bells are ringing between short pauses," and the right column is a continuous sound ``a small dog snoring and groaning". (a) and (b) are their spectrograms. (c) and (d) are their interpolated adjacency graphs of GraphAC. (e) and (f) are their interpolated adjacency graphs of GraphAC without top-{\it{k}} mask (GraphAC w/o top-{\it{k}}). The blue contours denote the meaningless areas with over attention, and the green contour denotes the important area with insufficient attention by the GraphAC w/o top-{\it{k}}.
}
	\vspace{-6mm}
	\label{fig:semantic}
\end{figure}

Fig.~\ref{fig:semantic} illustrates the mechanism of the graph attention for the audio feature representation, with two audio clips as examples. The clip in the left column is a transient sound of ``five different sounding bells are ringing between short pauses," and the clip in the right column is a continuous sound of ``a small dog snoring and groaning". Their spectrograms are shown in Fig.~\ref{fig:semantic}(a) and (b). The corresponding adjacency graphs $\hat{\mathbf{A}}$ learnt with the top-{\it{k}} mask are in Fig.~\ref{fig:semantic}(c) and (d), and those without the top-{\it{k}} mask are in Fig.~\ref{fig:semantic}(e) and (f), with bilinear interpolation applied for better illustration.

It can be found that the interpolated adjacency graph displays the vertical patterns. The most possible reason is that the graph attention mechanism employs an additive score function (i.e., Eq. \eqref{eq:relation_coefficient} normalised by softmax function) to obtain the attention coefficients between the audio nodes. It focuses on whether the nodes contain important information about the scenes and events, with greater attention coefficients implying more important nodes than others.

The interpolated adjacency graph highlights audio feature nodes with long-time dependency on acoustic scenes and events. As shown in~Fig.~\ref{fig:semantic}(c) and (d), it works for both transient and continuous sound. The highlighted audio feature nodes have a vertical bar pattern in the adjacency graph. This is because the graph modelling in GraphAC is applied on $\mathbf{x}_i, 1 \le\!i\le\!T$, i.e., the audio feature node at each time frame. The learnt node relations are reflected as the asymmetric directed adjacency graph $\hat{\mathbf{A}}$, which highlights time-dependencies between audio nodes. The selected audio feature nodes by the adjacency graph are those with long-time dependencies among all nodes that can help capture the contextual information from the audio signal, and thus the semantic information about acoustic scenes and events.

\subsection{Effect of the Top-{\it{k}} Mask}
We compared GraphAC and GraphAC without the top-{\it{k}} mask in graph attention (GraphAC w/o top-{\it{k}}) in Table \ref{tab:performance}. Results show that the performance without the top-{\it{k}} mask degrades in core semantic metrics, i.e., CIDE$_r$, SPICE and SPIDE$_r$. Examples of their adjacency graphs (bilinear interpolated) are shown in Fig. \ref{fig:semantic}(c)-(f). The adjacency graph generated by GraphAC w/o top-{\it{k}} has attention to the meaningless background audio feature nodes, contoured in blue in Fig. \ref{fig:semantic}(e) and (f), and insufficient attention to some important nodes, contoured in green in Fig. \ref{fig:semantic}(f). In contrast, the proposed method with the top-{\it{k}} mask can focus on the important audio feature nodes and ignore the meaningless audio feature nodes, when modelling the audio feature with long-time dependencies. In this work, $\textit{k}=25$ is selected empirically from the experiments. Future work will include the adaptive estimation of the $\textit{k}$ value from audio signals with different time duration.

\section{Conclusion}

We have presented a novel audio captioning method using graph attention in the encoder to exploit temporal dependencies in audio features. It facilitates the decoder in generating better captions with the timing and contextual information of the audio signal. Experiments show that the proposed method achieves state-of-the-art captioning performance. In addition, the proposed graph modelling enables audio feature representation with temporal information, which may benefit other tasks such as audio scene classification and event detection. Future work will investigate the latent relationship between audio feature nodes and the caption words by the graph learning, and develop methods for estimating $\textit{k}$ in the top-{\it{k}} mask adaptively to suit audio objects with different time spans.




\end{document}